\documentclass{PoS}
\usepackage{color,natbib}

\newcommand\degree{$^{\circ}\,$}

\newcommand\simlt{\lower.5ex\hbox{$\; \buildrel < \over \sim \;$}}
\newcommand\simgt{\lower.5ex\hbox{$\; \buildrel > \over \sim \;$}}

\title{Can we reconcile the TA excess and hotspot with Auger observations?}

\ShortTitle{TA vs Auger observations}

\author{Noemie Globus\\
        Racah Institute of Physics, The Hebrew University, 91904 Jerusalem, Israel\\
        E-mail: \email{noemie.globus@mail.huji.ac.il}}

\author{Denis Allard\\
        Laboratoire Astroparticule et Cosmologie, Universit\'e Paris Diderot/CNRS, 10 rue A. Domon et L. Duquet, F-75205 Paris Cedex 13, France\\
        E-mail: \email{allard@apc.in2p3.fr}}
        
\author{\speaker{Etienne Parizot}\\
        Laboratoire Astroparticule et Cosmologie, Universit\'e Paris Diderot/CNRS, 10 rue A. Domon et L. Duquet, F-75205 Paris Cedex 13, France\\
        E-mail: \email{parizot@apc.in2p3.fr}}
        
\author{Cyril Lachaud\\
        Laboratoire Astroparticule et Cosmologie, Universit\'e Paris Diderot/CNRS, 10 rue A. Domon et L. Duquet, F-75205 Paris Cedex 13, France\\
        E-mail: \email{lachaud@apc.in2p3.fr}}        
               
\author{Tsvi Piran\\
        Racah Institute of Physics, The Hebrew University, 91904 Jerusalem, Israel\\
        E-mail: \email{tsvi.piran@mail.huji.ac.il}}         

\abstract{The Telescope Array (TA) shows  a 20$^{\circ}$ hotspot as well as an excess of UHECRs above  50~EeV when compared with the Auger spectrum. We consider the possibility that both the TA excess and  hotspot are due to  a dominant source in the Northern sky. We carry out detailed simulations of UHECR propagation in both the intergalactic medium and the Galaxy, using different values for the intergalactic magnetic field. 
We  consider two general classes of sources: transients and steady, adopting  a mixed UHECR composition that is consistent with the one found by Auger. The spatial location of the sources is draw randomly.
We generate Auger-like and TA-like data sets from which we determine the spectrum, the  sky maps  and  the level of anisotropy.
We find that, while steady sources are favored over transients,  it is unlikely to account for all the currently available observational data. 
Most of the simulated data sets with a flux excess compatible with TA (at most a few percent depending on density model) show a much stronger anisotropy than the one observed. 
We find that the rare cases in which both the spectrum and the anisotropy are consistent require a steady source within $\sim 10$ Mpc, to account for the flux excess, and a strong extragalactic magnetic field $\sim 10$ nG, to reduce the excessive anisotropy. }

\FullConference{35th International Cosmic Ray Conference --- ICRC2017\\
		10--20 July, 2017\\
		Bexco, Busan, Korea}

\begin{document}

\section{Introduction}
The possible difference in composition of the UHECRs observed by TA and by Auger has been widely discussed, after the claim by TA that their data is compatible with a pure proton composition (see for instance \cite{TACompo}, whereas Auger reports a gradual, but very significant trend towards higher mass nuclei around 10~EeV \cite{AugerCompo1,AugerCompo2}. 
A joint analysis \citep{Unger15} showed that the TA data is not inconsistent with the transition towards heavier elements inferred from the Auger data\footnote{see however \citet{ShahamPiran13} for a different interpretation of the evolution of the composition.}.

The difference in the clustering of events 
appears more significant. While no significant small or intermediate-scale anisotropy can be observed in the Auger data [the largest departure from isotropy was found to have a post-trial probability of $\sim1.4$\% \citep{Aab15} in the direction of Cen A], the TA Collaboration reported a so-called hotspot, with a $20^{\circ}$ angular scale, near the constellation Ursa Major.
The chance probability of observing such a clustering anywhere in the sky is $3.7\cdot10^{-4}$, equivalent to a one-sided probability of $3.4~\sigma$ \citep{Tinyakov15}. { The highest energy events are not present in the hotspot region itself. This can be explained by simply noting that the rigidity of the highest energy particles is smaller than those at intermediate energies, due to the change in composition.}

\begin{figure}[!h]
\begin{center}
\includegraphics[width=0.6\textwidth]{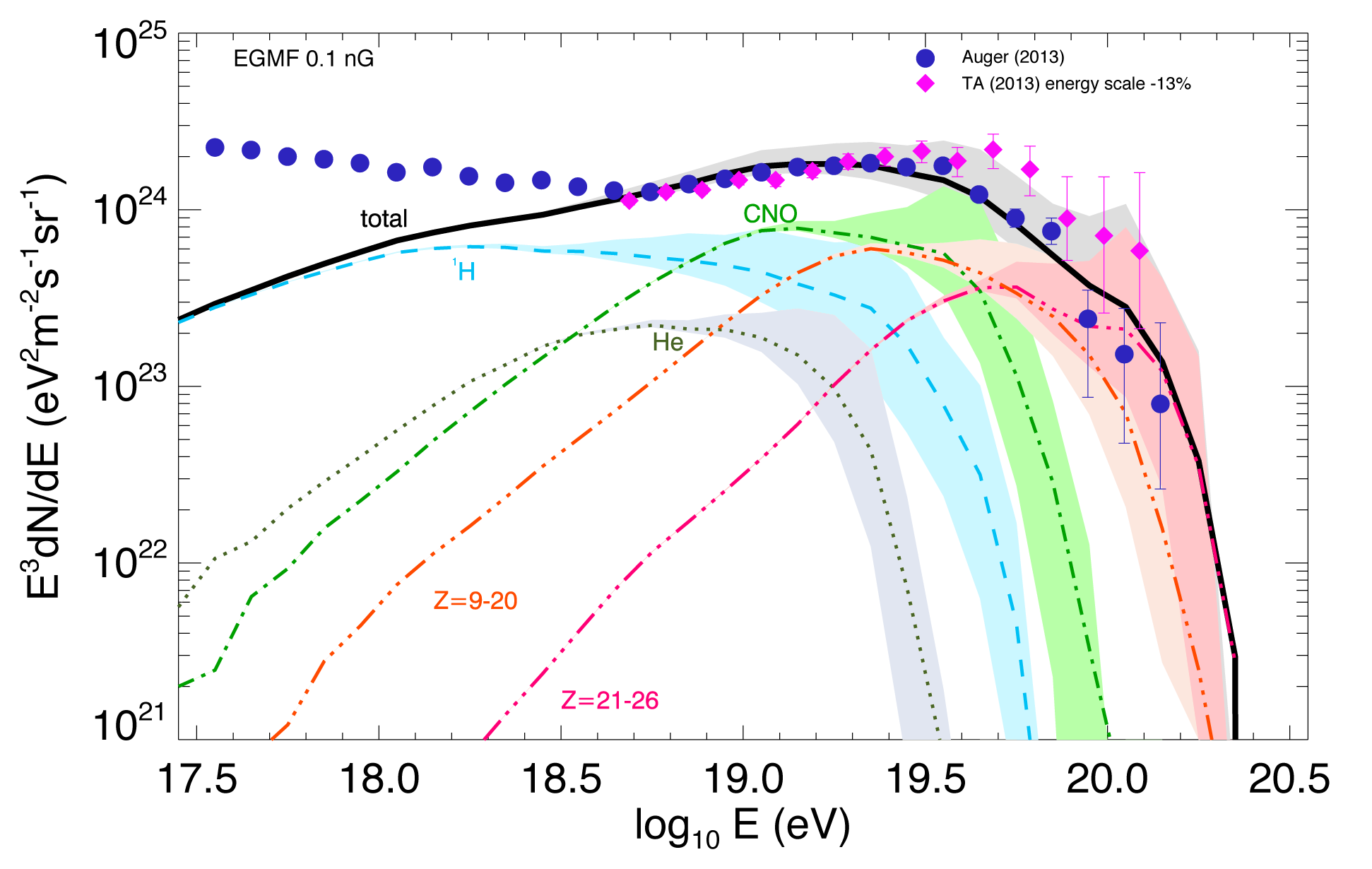}
\caption{Diffuse cosmic ray flux spectrum, expected on Earth, assuming $B_{\mathrm{EGMF}} = 0.1$~nG, from \citet{Globus15a}. The contributions of different groups of nuclei are shown, the lines (plain lines for the total spectrum) represent the mean value calculated over 300 realizations of a transient source scenario (here according to the GRB rate), the shaded areas represent the 90\% intervals (excluding the 5\% highest and the 5\% lowest realizations) of the 300 realizations.  These fluxes are compared with the latest Auger and TA estimates of the UHECR flux (a shift of $-13\%$ has been applied to the TA energy scale).
 }
\label{fig:model}
\end{center}
\end{figure}

While such a level of significance is too low to be conclusive, it should be considered together with an other difference, regarding the energy spectrum above $\sim50$~EeV.
Fig.~\ref{fig:model} depicts the Auger and TA data, where a shift of $-13\%$ has been applied to the TA energy scale, as recommended by the Auger-TA joint working group \citep{Unger15}. The TA spectrum clearly shows a significant excess at higher energy, at least if one considers only the statistical error bars (shown on the plot). A systematic uncertainty with a rather strong energy dependence would be needed to explain such a difference.

After scaling down the energy by 13\% there are 83 highest energy TA events above 50~EeV.
They correspond to an exposure of 8,600~km$^{2}\,\mathrm{sr}\,\mathrm{yr}$ \citep{matthews_talk}. On the other hand, Auger reports 231 events above 52~EeV, for an exposure of 66,452~km$^{2}\,\mathrm{sr}\,\mathrm{yr}$. Given the shape of the spectrum between 50~and 60~EeV, this extrapolates to  $\sim 290$ events above 50~EeV. If the Auger flux is assumed to represent the average UHECR flux in the absence of anisotropy, then the expected number of events for TA is $\sim 38$. The actual integrated flux of TA would thus need to be a 7$\sigma$ upward fluctuation.

It thus appears unlikely that the UHECR fluxes observed by Auger and TA are just different realisations of an underlying roughly isotropic flux. 
Put together with the observation of the TA hotspot, the current data suggest 
that this excess is caused by the contribution of one (or more) localised sources in the Northern sky. Quantitatively, if the integrated flux of Auger above 50~EeV represents an average contribution of typical sources distributed more or less isotropically over the sky, the corresponding contribution in the TA data should be $\sim 38\pm 6$~events, which leaves $\sim 45\pm 6$ for the putative additional source(s). Thus, if the difference between the two spectra is attributed to a dominant source, this source should contribute 45\%--60\% of the total Northern sky flux.

\section{Transient vs permanent sources: analytical estimates}
{
The probability for a transient source (occuring at a rate $R_{-9}10^{-9}$ Mpc$^{-3}$ yr$^{-1}$)  to contribute a fraction $\eta_{\mathrm{flux}}$ of the total UHECR flux is \citep{Globus16}
\begin{eqnarray}
\mathcal{P}(\eta_{\mathrm{flux}}) \simeq& 4.3\% \,\, \eta_{\mathrm{flux}}^{-5/4} \, E_{20}^{1/2} (Z B_{\mathrm{nG}})^{-1/2}  \, R_{-9}^{-1/4} \, \lambda_{\mathrm{Mpc}}^{-1/4}\, H_{100}^{-5/4}
\label{eq:probaEtaFluxBurst}
\end{eqnarray}
where $E_{20} 10^{20}$ eV is the energy of the cosmic-ray nuclei of charge $Z$, $B_{\mathrm{nG}}$ the strength of the extragalactic magnetic field (EGMF) in nG, $\lambda_{\mathrm{Mpc}}$ the EGMF coherence length in Mpc and  $H_{100} 100$ Mpc the GZK horizon. We assumed here that the sources are standard candles.
The angular size of such a source in the sky is
\begin{equation}
\Delta\theta(\eta_{\mathrm{flux}})\simeq 1.3^{\circ} \, \eta_{\mathrm{flux}}^{-1/8} \, E_{20}^{-3/4}\, (ZB_{\mathrm{nG}})^{3/4} \lambda_{\mathrm{Mpc}}^{3/8} \, (R_{-9}H_{100})^{-1/8}.\nonumber\\ 
\label{eq:angularSize}
\end{equation}

With this simple estimates, we see that the probability for a transient source of CNO at 50 EeV ($H_{100}\sim2$) to contribute 50\% of the total flux, for  $R_{-9}=B_{\mathrm{nG}}=\lambda_{\mathrm{Mpc}}=1$ is $\sim~1$\%. The source would span over 9\degree in the sky. The Galactic magnetic field (GMF) would induce additional deflections. Its effect would be taken into account in our numerical simulations.\\
For steady sources, the probability to find a source does not depends on the value of the magnetic field, only on the source density $n_{-5}\,10^{-5}$ Mpc$^{-3}$, and it is given by
\begin{equation}
\mathcal{P}(\eta_{\mathrm{flux}}) \simeq 3.0\% \, \eta_{\mathrm{flux}}^{-3/2} \, n_{-5}^{-1/2} \, H_{100}^{-3/2}.
\label{eq:probaEtaFluxSteady}
\end{equation}

Regarding the angular spread of such a source, it is estimated as
\begin{equation}
\Delta\theta(\eta_{\mathrm{flux}}) \simeq 1.1^{\circ} \,\, \eta_{\mathrm{flux}}^{-1/4} \, Z E_{20}^{-1} \, B_{\mathrm{nG}} \,\,\lambda_{\mathrm{Mpc}}^{1/2} \,\, n_{-5}^{-1/4} \, H_{100}^{-3/4}.
\label{eq:angularSizeSteady}
\end{equation}
Transient sources suffer from a general problem: larger deflections also imply larger spreads in the particles arrival time, which in turn reduce the apparent flux of the source, and thus makes it even less likely for a source to contribute a large fraction of the total UHECR flux. Steady sources, on the other hand, do not suffer from this problem, since their apparent flux does not depend on the time spread, but only on their distance. Larger magnetic fields, at least in the direction of the source, might thus 
 increase its apparent angular size, without reducing its flux. 
}

\section{Model}

A model must provide different spectra in the Northern and Southern hemispheres and also reproduce the anisotropy patterns: it must i) provide a hotspot in the Northern sky with a typical angular scale of 20$^{\circ}$, ii) be compatible with isotropy in the Southern sky (i.e not produce an anisotropy signal much stronger than the warm spot reported around the direction of Cen A).

Concerning the other observable of UHECR phenomenology, namely the composition, it is taken into account here in a generic way. In \citet{Globus15a}, some of us have developed a model based on the acceleration of particles in the mildly relativistic internal shocks of gamma-ray bursts (GRBs). This model  reproduces the spectrum and composition 
both below and above the ankle \citep{Globus15b}. From a phenomenological point of view, the main features of this model are a low value of the maximum energy for protons at the sources, a hard source spectrum for all nuclei except protons (which have a significantly softer spectrum), and a source composition with a metallicity higher than the usual Galactic cosmic-ray component by a factor of $\sim 10$. These are considered here as generic features of a working model, providing a suitable description of the average UHECRs, independently of the actual sources, whether GRBs, other types of transient sources \citep[like tidal disruption events, see e.g.][]{FG09,FP14,Komossa15}, or steady sources. For the purpose of the anisotropy analyses of this paper, the main relevant ingredient is the composition of the UHECRs with an energy larger than 50~EeV, which is thus assumed to be the same as that of our explicit GRB model \citep{Globus15a}, but without prejudice regarding the nature of the sources.

\section{Simulation procedure}
We created data sets adapted to Auger and TA current statistics
, sky coverage and energy resolution, in the case of transient and steady sources (with different source densities), and analyzed them from the point of view of their flux excess in the TA sky and their anisotropy. The data sets were built, for each realisation, with the same number of cosmic-ray events as the data to which it is compared (\textit{i.e.} with the 83 highest energy events in the case of the TA, and with the 231 highest energy events in the case of Auger). 
{ Our detailed simulations took into account the various effects influencing the propagation of the UHECRs, including energy losses, photodissociation in the case of nuclei, and deflections by the intervening magnetic fields, around the source, in the intergalactic medium and in the Galaxy.
We considered a purely turbulent EGMF with a Kolmogorov spectrum and a coherence length of 200 kpc. 
For the GMF we used the \citet{JF12} model. We tested two different coherence length for the turbulent component of the GMF \citep[50 and 200 pc according to][] {Beck16}.}
The free parameters of the models are the EGMF value, the GMF coherence length and the source density or occurence rate. 
We produced 1,200 random realisations of the source distribution in the case of transient sources \citep[according to the GRB occurrence rate $1.3\,10^{-9}\,\mathrm{Mpc}^{-3}\mathrm{yr}^{-1}$, see][]{WP10}, and 600 random realisations of the  source distribution in the universe in the case of steady sources, with source densities: $10^{-4}\,\rm{Mpc}^{-3}$ and $10^{-5}\,\rm{Mpc}^{-3}$.
{ For each of these realisations, we produced 10 random data sets whose differences reflect statistical fluctuations} \citep[more details in][]{Globus16}.
\begin{figure}[!h]
\begin{center}
\includegraphics[width=0.65\textwidth]{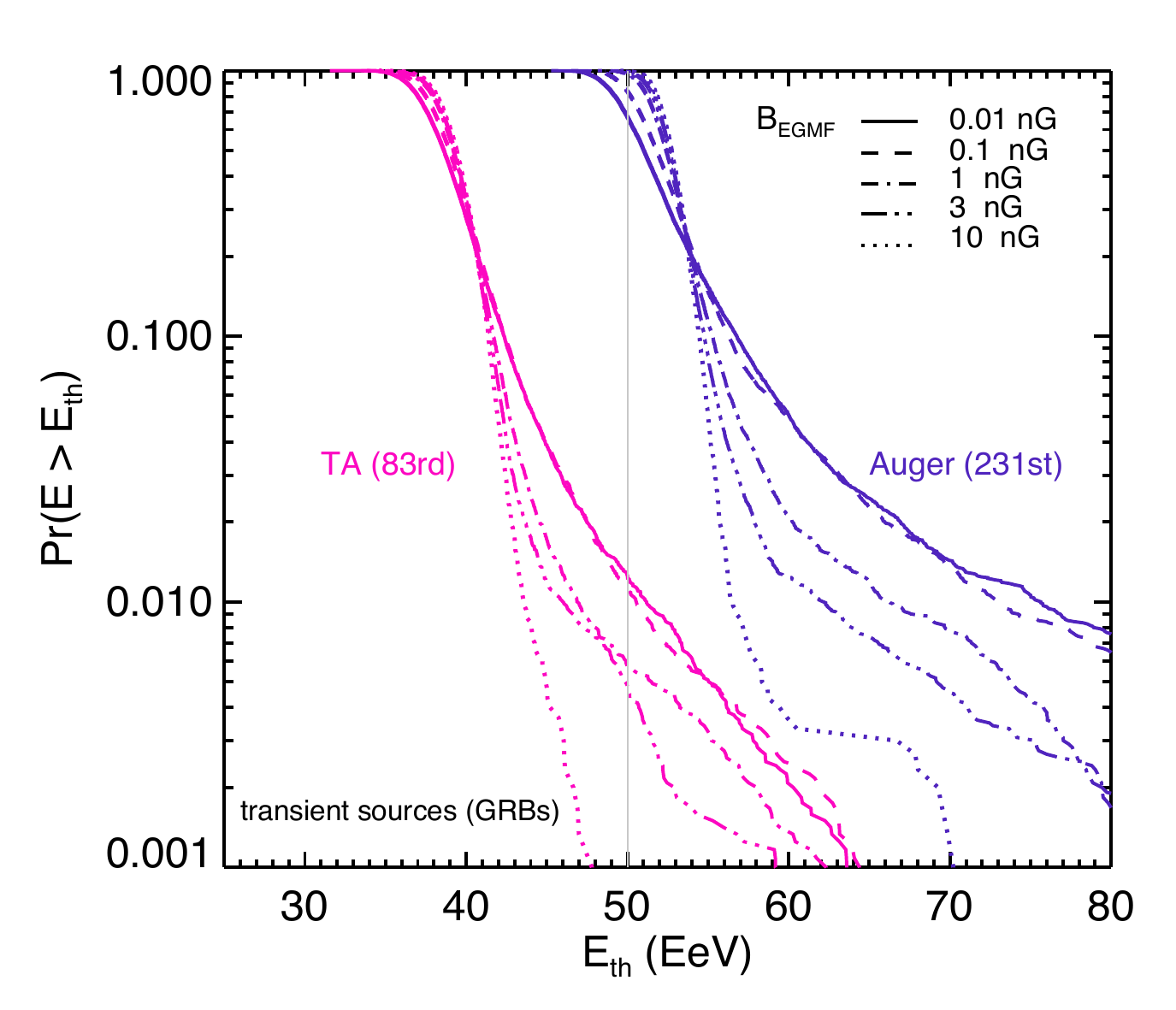}\\\includegraphics[width=0.65\textwidth]{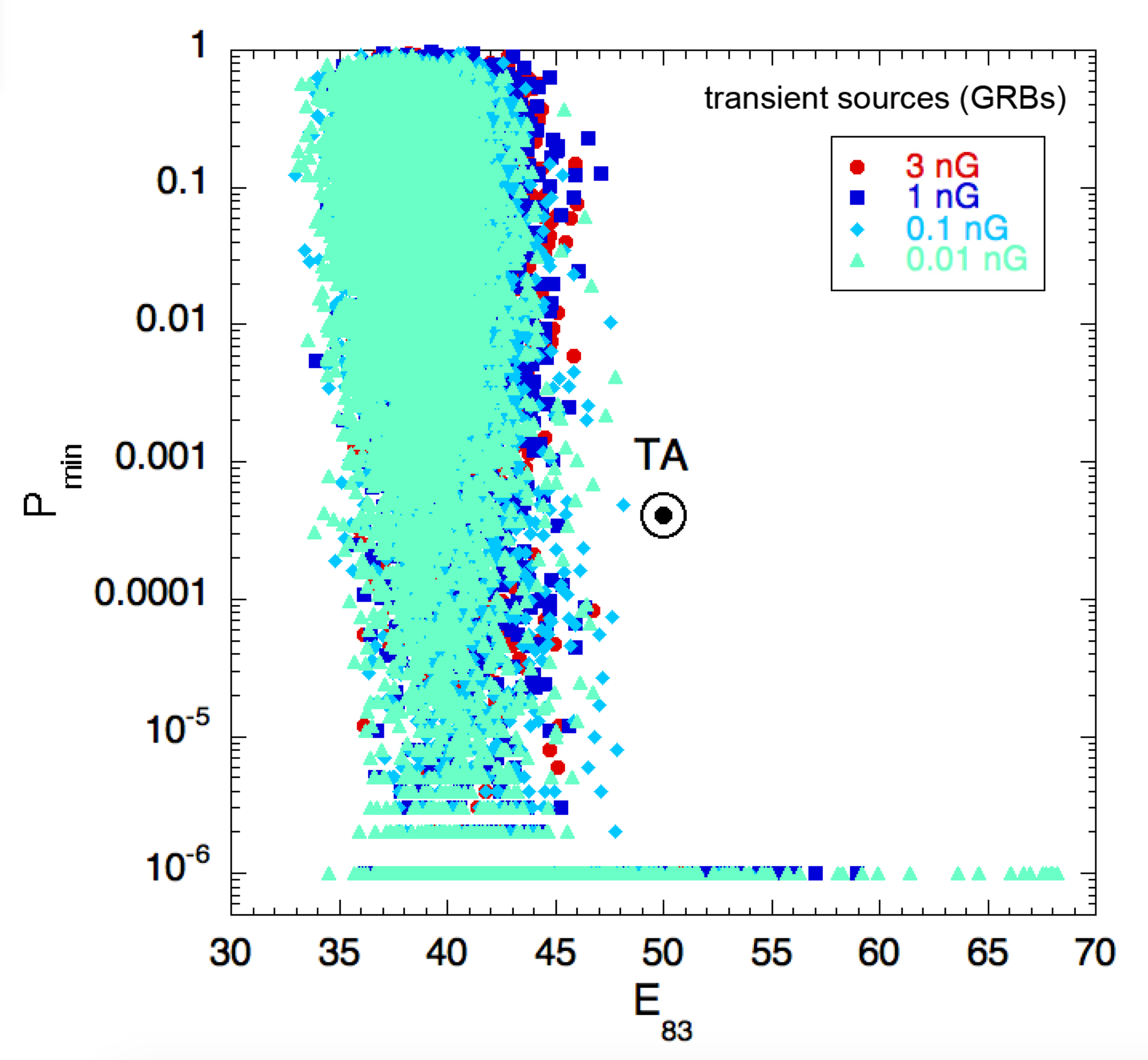}
\caption{Upper panel:
Cumulative probability distribution of the energy of the 83$\mathrm{rd}$ highest energy events in the simulated TA-like data sets, $E_{83}$ (pink), and of the 231$^{\mathrm{st}}$ highest energy events in the simulated Auger-like data sets, $E_{231}$ (violet), for the transient source (GRB) model with different values of the EGMF, as indicated. Shown is the probability that $E_{83}$ and $E_{231}$ are larger than the energy given in abscissa. Lower panel:
A scatter plot of the values of $\mathcal{P}_{\min}$ vs. $E_{83}$ for all the TA-like data sets simulated in the transient source model, with 4 values of the EGMF, as indicated. The position of the actual TA data set is indicated by the $\odot$ symbol. 
Note that only the realisations fulfilling both the Auger 2-point and Auger flux criteria  are considered in this scatter plot. The figures are taken from \citet{Globus16}.
 }
\label{fig:transient}
\end{center}
\end{figure}

\begin{figure}[!h]
\begin{center}
\includegraphics[width=0.65\textwidth]{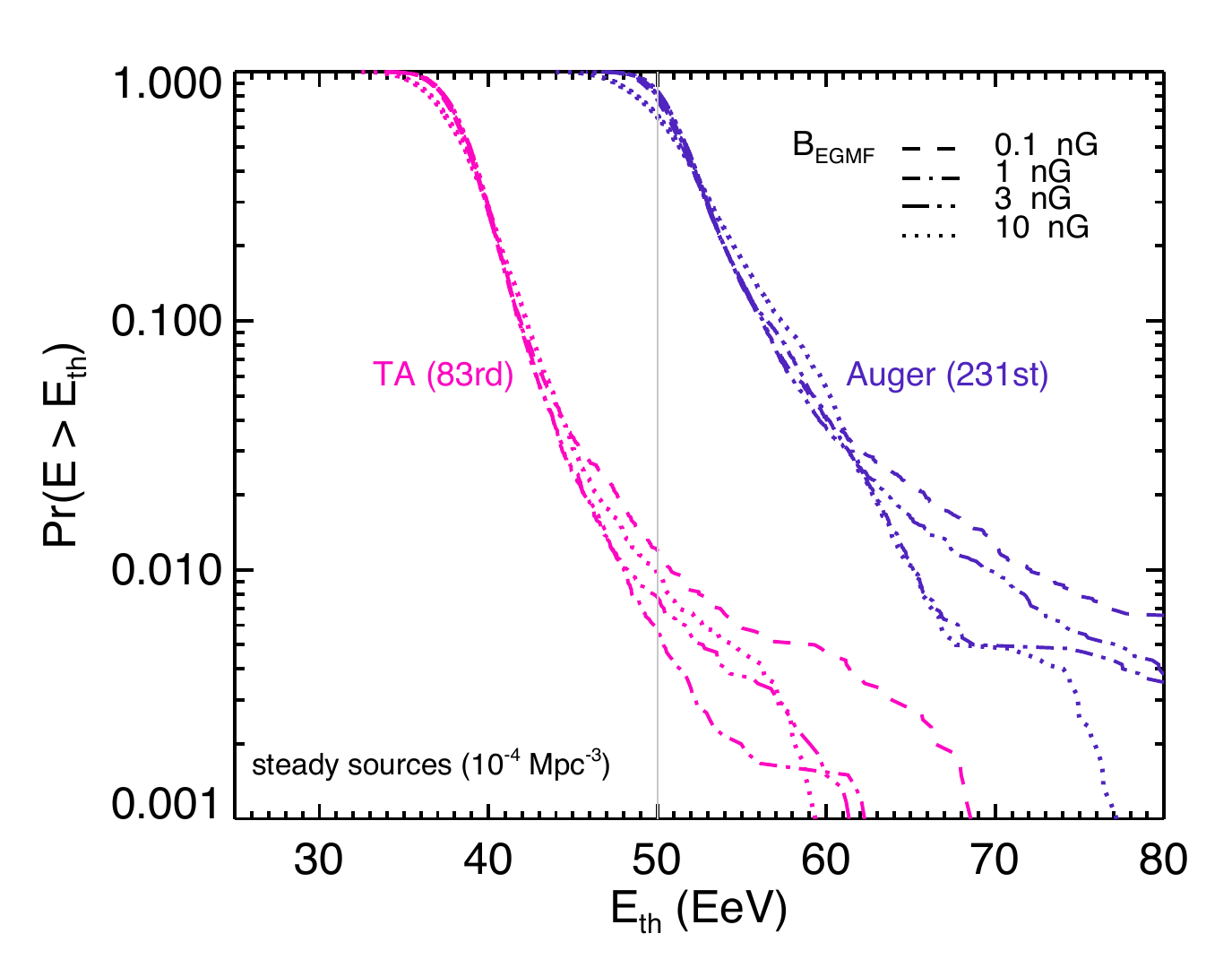}\\\includegraphics[width=0.65\textwidth]{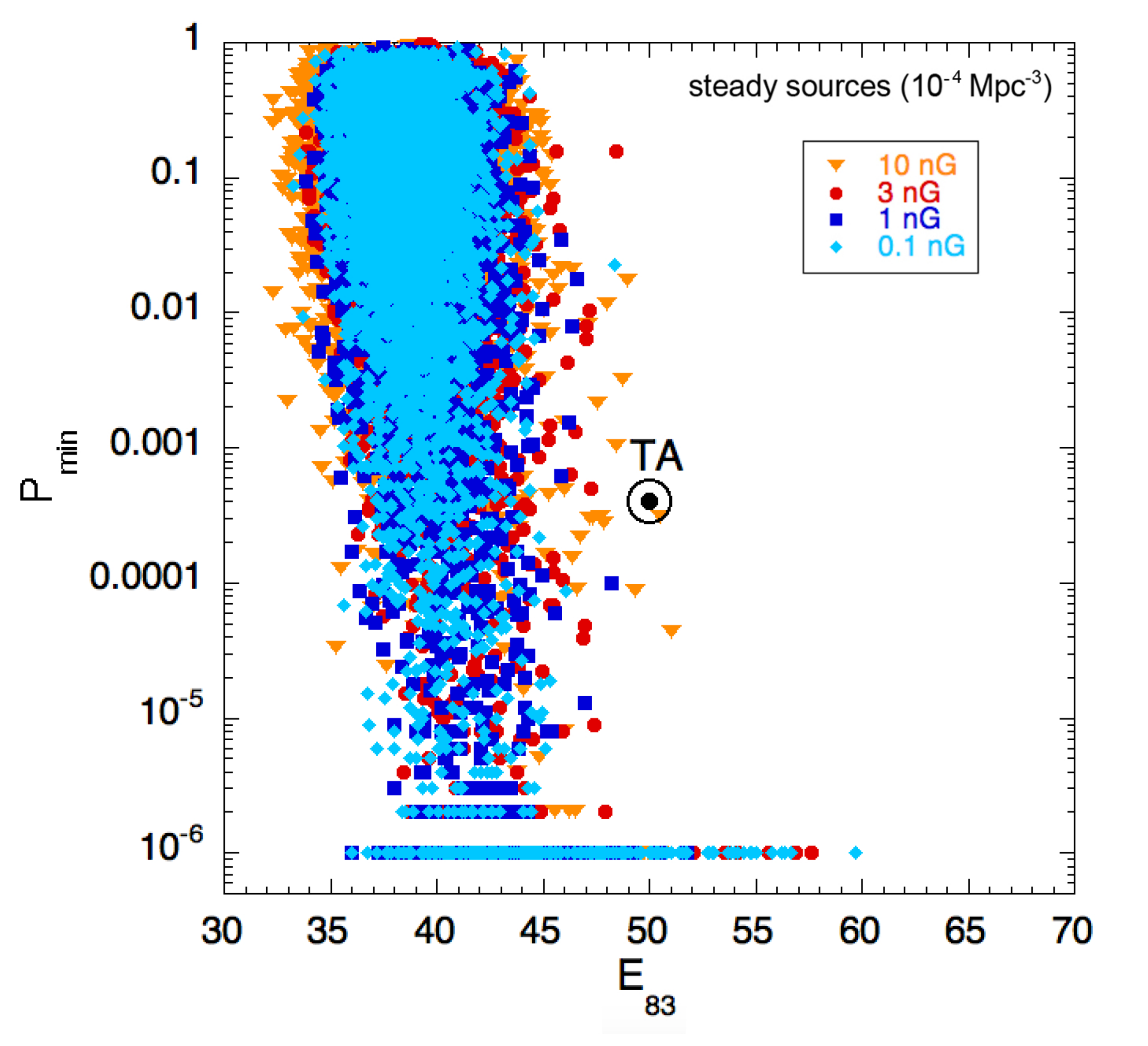}
\caption{Same as Figure \ref{fig:transient}, but in the steady source model. The source density is $10^{-4}\,\rm{Mpc}^{-3}$. The figures are taken from \citet{Globus16}.}
\label{fig:steady}
\end{center}
\end{figure}

\section{Results}

We first considered the flux excess in the Northern sky. We asked what is the energy of the 83$^{\mathrm{rd}}$ most energetic event, namely $E_{83}$~EeV.
The upper panel of Fig.~\ref{fig:transient} shows the cumulative probability distribution of $E_{83}$ among all realisations of the transient source model 
for four different values of the extragalactic magnetic field variance. 
As expected, the value of $E_{83}$ is usually much lower than the actual value for TA (50~EeV after rescaling the TA energy scale). 
For $B_{\mathrm{EGMF}}=10$~nG, it is extremely improbable that a source can have a large enough contribution in the TA sky to explain the observed excess. 
For $B_{\mathrm{EGMF}} < 0.1$~nG, the occurrence rate may reach around 1\%. 
On the same figure, we also plot the value of $E_{231}$ for the Auger-like simulated data sets. The actual value of $E_{231} = 52$~EeV is quite common.

To study the anisotropy, we  performed a 2-point correlation function analysis\footnote{we also performed a clustering analysis, see \citet{Globus16} for more details. }, \textit{i.e.}  
we calculated the probability  $\mathcal{P}_{\mathrm{iso}}(\theta)$ that a purely isotropic UHECR flux would produce at least as many pairs of events separated by an angle lower than $\theta$. 
In the actual TA data set, the value smallest value, $\mathcal{P}_{\min} \simeq 4\cdot10^{-4}$, is reached at the angular scale $\theta_{\min}\simeq 25^{\circ}$ \citep{Tinyakov15}. 

Since our aim was to account for both Auger and TA data, we first implemented a cut rejecting realisations showing a flux excess and/or strong anisotropies in the Auger sky. We then studied the correlation between the flux excess and anisotropy in the TA sky. The correlation between  $\mathcal{P}_{\min}$ and $E_{83}$ is shown in the lower panel of Fig.~\ref{fig:transient} in the case of transient sources. All the TA data sets with $E_{83}> 50$ EeV have $\mathcal{P}_{\min}\leq10^{-6}$. This means that those data sets show very significant anisotropies at small angular scales\footnote{in order to evaluate $\mathcal{P}_{\min}$, we computed one million random realisations of an isotropic flux. Therefore, we cannot attribute values to $\mathcal{P}_{\min}$ lower than $10^{-6}$.}, much more anisotropic than what is observed. 

Results for the steady source scenario are shown in Fig. \ref{fig:steady} for a source density of $10^{-4}\,\rm{Mpc}^{-3}$. As predicted, the situation is slightly better than in the transient source scenario: few data sets yielding a strong flux excess ($E_{83}\geq50$ EeV) in the largest EGMF case do not necessarily show prohibitive values of $\mathcal{P}_{\min}$ and lie in the vicinity of the TA data point.\\

To summarize, we have addressed the compatibility between the Auger and Telescope Array data, in the framework of an extragalactic UHECR source model, and found that:\\
- The flux excess in the TA sky could typically occur at most in a few percent of the cases, either in the transient source scenario or in the steady source scenario.\\
- The conjunction between a large flux excess and a moderate anisotropy in the Northern hemisphere turned out to be particularly challenging to account for. We find that transient sources are essentially incapable of reproducing the data. 
We find that the rare 
cases in which both the spectrum and the anisotropy are consistent require a steady source within $\sim10$ Mpc, to account for the flux excess, and a strong extragalactic magnetic field $B_{\mathrm{EGMF}} \geq10$~nG, to reduce the excessive anisotropy.


\section*{Acknowledgements}
This research was supported by I-Core CHE-ISF center of excellence of research in astrophysics (NG \& TP), by the Lady Davis foundation (NG), and by the UnivEarthS Labex program at Sorbonne Paris Cit\'e (ANR-10-LABX-0023 and ANR-11-IDEX-0005-02) (CL).

\end{document}